\def\rn{\noindent\parshape 2 0truecm 8.8truecm 0.3truecm 8.5truecm}
\def\nn#1 #2{#1, #2.}				
\def\nnn#1 #2 #3{#1, #2. #3.}			
\def\nnnn#1 #2 #3 #4{#1, #2. #3. #4.}		
\def\nnnnn#1 #2 #3 #4 #5{#1, #2. #3. #4. #5.}	
\def\dualand{, \&\hbox{ }}				
\def\multiand{, \&\hbox{ }}				

\def\rg#1;#2;#3;#4;#5;#6 {\par\rn#1 #2, {\it #3}, {\bf #4}, #5 (``#6'') \par}
\def\rf#1;#2;#3;#4;#5 {\par\rn#1 #2, {\it #3}, {\bf #4}, #5\par}
\def\rfbook#1;#2;#3;#4;#5 {{\frenchspacing\par\rn#1 #2, {\it #3} (#4: #5)\par}}
\def\rfproc#1;#2;#3;#4;#5;#6 {{\frenchspacing\par\rn#1 #2, in {\it #3}, ed. #4 (#5: #6)\par}}
\def\rfprep#1;#2;#3  {{\par\rn#1 #2, #3\par}}
\def\rfprepp#1;#2;#3 {{\par\rn#1 #2, #3\par}}

\def\K{{\rm K}}

\def\expec#1{\langle#1\rangle}

\def\etal{{\frenchspacing\it et al.}}
\def\ie{{\frenchspacing\it i.e.}}
\def\eg{{\frenchspacing\it e.g.}}
\def\etc{{\frenchspacing\it etc.}}

\def\beq#1{\begin{equation}\label{#1}}
\def\eeq{\end{equation}}
\def\beqa#1{\begin{eqnarray}\label{#1}}
\def\eeqa{\end{eqnarray}}
\def\eq#1{equation~(\ref{#1})}

\def\fig#1{Figure~\ref{#1}}
\def\Fig#1{Figure~\ref{#1}}

\def\sec#1{Section~\ref{#1}}

\def\spose#1{\hbox to 0pt{#1\hss}}
\def\simlt{\mathrel{\spose{\lower 3pt\hbox{$\mathchar"218$}}
     \raise 2.0pt\hbox{$\mathchar"13C$}}}
\def\simgt{\mathrel{\spose{\lower 3pt\hbox{$\mathchar"218$}}
     \raise 2.0pt\hbox{$\mathchar"13E$}}}
\def\simpropto{\mathrel{\spose{\lower 3pt\hbox{$\mathchar"218$}}
     \raise 2.0pt\hbox{$\propto$}}}

\def\ed{\end{document}}

\def\Ob{\Omega_{\rm b}}
\def\Oc{\Omega_{\rm cdm}}
\def\Ok{\Omega_{\rm k}}
\def\Ol{\Omega_\Lambda}
\def\Om{\Omega_{\rm m}}
\def\On{\Omega_\nu}
\def\ob{\omega_{\rm b}}
\def\oc{\omega_{\rm cdm}}

\def\om{\omega_{\rm m}}
\def\on{\omega_\nu}
\def\Cl{C_\l}
\def\dT{\delta T}
\def\T{{\bf d}}
\def\ns{n_s}
\def\nt{n_t}
\def\As{A_s}
\def\At{A_t}
\def\dA{d_{\rm lss}}

\def\data{{\rm data}}
\def\L{{\cal L}}

\def\p{{\bf p}}
\def\x{{\bf x}}

\def\C{{\bf C}}
\def\I{{\bf I}}
\def\M{{\bf M}}

\def\l{\ell}
\def\llo{\l_{\rm low}}
\def\lhi{\l_{\rm high}}
\def\Cl{C_\ell}

\def\lstar{\l^*}

\documentstyle[emulateapj,danonecolfloat]{article}

\begin{document}
\twocolumn[


\journalid{337}{15 January 1989}
\articleid{11}{14}

\submitted{Submitted to ApJ February 11 2000, accepted April 6}

\title{Current cosmological constraints from a 10 parameter CMB analysis}

\author{
Max Tegmark
\footnote{Dept. of Physics, Univ. of Pennsylvania, 
Philadelphia, PA 19104;
max@physics.upenn.edu}$^{,\>b}$
and 
Matias Zaldarriaga
\footnote{Institute for Advanced Study, Princeton, 
NJ 08540; matiasz@ias.edu}
}

\begin{abstract}

We compute the constraints on a ``standard'' 10 parameter 
cold dark matter (CDM) model from the most recent 
CMB data and other observations, 
exploring 30 million discrete models and 
two continuous parameters.
Our parameters are the densities of CDM, baryons, neutrinos, 
vacuum energy and curvature, the reionization optical depth, and the
normalization and tilt for both scalar and tensor fluctuations.
Our strongest constraints are on spatial curvature,
$-0.24<\Ok<0.38$, and CDM density,
$h^2\Oc<0.3$, both at 95\%.
Including SN 1a constraints gives a
positive cosmological constant at high significance.
We explore the robustness of our results to various
assumptions. We find that three different data subsets 
give qualitatively consistent constraints.
Some of the technical issues that have the largest impact
are the inclusion of calibration errors, closed models,
gravity waves, reionization, nucleosynthesis constraints 
and 10-dimensional likelihood interpolation.

\end{abstract}


]


\section{INTRODUCTION}

The past year has yet again seen dramatically improved measurements
of the Cosmic Microwave Background (CMB) power spectrum, 
with the Python, Viper, Toco and Boomerang experiments suggesting 
a first acoustic peak with a fairly well-defined height and position.
Further great improvements are expected shortly from the Antarctic 
Boomerang flight, the MAP satellite and other experiments, with the potential 
to accurately measure about ten cosmological parameters
(Jungman {\etal} 1996; 
Bond {\etal} 1997; Zaldarriaga {\etal} 1997;
Efstathiou \& Bond 1998),
especially when combined with galaxy redshift surveys
(Eisenstein {\etal} 1999), supernovae 1a (SN 1a) 
observations (White 1998) or gravitational Lensing
(Hu \& Tegmark 1999).
\begin{table}
\noindent
{
\small
{\bf Table 1} -- CMB data used\\
}
\smallskip
{\footnotesize
\begin{tabular}{|l|r|r|}
\hline
Experiment	&$\dT$	&$\l$\\
\hline
COBE            &$  8.5^{+16.0}_{- 8.5}$  &$  2.1^{+  0.4}_{-  0.1}$\\
COBE            &$ 28.0^{+ 7.5}_{-10.3}$  &$  3.1^{+  0.6}_{-  0.6}$\\
COBE            &$ 34.0^{+ 6.0}_{- 7.2}$  &$  4.1^{+  0.7}_{-  0.7}$\\
COBE            &$ 25.1^{+ 5.3}_{- 6.6}$  &$  5.6^{+  1.0}_{-  0.9}$\\
COBE            &$ 29.4^{+ 3.6}_{- 4.1}$  &$  8.0^{+  1.3}_{-  1.2}$\\
COBE            &$ 27.7^{+ 3.9}_{- 4.5}$  &$ 10.9^{+  1.3}_{-  1.2}$\\
COBE            &$ 26.1^{+ 4.4}_{- 5.2}$  &$ 14.4^{+  1.3}_{-  1.6}$\\
COBE            &$ 33.0^{+ 4.6}_{- 5.4}$  &$ 19.4^{+  2.7}_{-  2.8}$\\
FIRS            &$ 29.4^{+ 7.8}_{- 7.7}$  &$ 11.0^{+ 17.0}_{-  9.0}$\\
Tenerife        &$ 32.5^{+10.1}_{- 8.5}$  &$ 20.0^{+ 10.0}_{-  8.0}$\\
IACB            &$111.9^{+65.4}_{-60.1}$  &$ 33.0^{+ 26.0}_{- 16.0}$\\
IACB            &$ 54.6^{+27.2}_{-21.9}$  &$ 53.0^{+ 26.0}_{- 19.0}$\\
SP              &$ 30.2^{+ 8.9}_{- 5.5}$  &$ 61.0^{+ 41.0}_{- 31.0}$\\
SP              &$ 36.3^{+13.6}_{- 6.1}$  &$ 61.0^{+ 41.0}_{- 31.0}$\\
BAM             &$ 55.6^{+29.6}_{-15.2}$  &$ 74.0^{+ 82.0}_{- 47.0}$\\
Python          &$ 60.0^{+15.0}_{-13.0}$  &$ 88.0^{+ 17.0}_{- 39.0}$\\
Python          &$ 66.0^{+17.0}_{-16.0}$  &$170.0^{+ 69.0}_{- 50.0}$\\
ARGO            &$ 39.1^{+ 8.7}_{- 8.7}$  &$ 95.0^{+ 78.0}_{- 44.0}$\\
ARGO            &$ 46.8^{+ 9.5}_{-12.1}$  &$ 95.0^{+ 78.0}_{- 44.0}$\\
IAB             &$ 94.5^{+41.8}_{-41.8}$  &$120.0^{+101.0}_{- 55.0}$\\
MAX             &$ 49.4^{+ 7.8}_{- 7.8}$  &$139.0^{+108.0}_{- 67.0}$\\
Saskatoon       &$ 49.0^{+ 8.0}_{- 5.0}$  &$ 87.0^{+ 44.0}_{- 35.0}$\\
Saskatoon       &$ 69.0^{+ 7.0}_{- 6.0}$  &$166.0^{+ 39.0}_{- 48.0}$\\
Saskatoon       &$ 85.0^{+10.0}_{- 8.0}$  &$237.0^{+ 36.0}_{- 48.0}$\\
Saskatoon       &$ 86.0^{+12.0}_{-10.0}$  &$286.0^{+ 33.0}_{- 44.0}$\\
Saskatoon       &$ 69.0^{+19.0}_{-28.0}$  &$349.0^{+ 51.0}_{- 46.0}$\\
CAT             &$ 50.8^{+15.4}_{-15.4}$  &$397.0^{+ 84.0}_{- 65.0}$\\
CAT             &$ 49.0^{+19.1}_{-13.6}$  &$615.0^{+102.0}_{- 72.0}$\\
CAT             &$ 54.0^{+ 9.5}_{- 6.4}$  &$397.0^{+ 84.0}_{- 65.0}$\\
CAT             &$ 43.6^{+13.6}_{-13.1}$  &$615.0^{+102.0}_{- 72.0}$\\
OVRO            &$ 56.0^{+ 8.5}_{- 6.6}$  &$537.0^{+267.0}_{-205.0}$\\
QMAP            &$ 47.0^{+ 6.0}_{- 7.0}$  &$ 80.0^{+ 41.0}_{- 41.0}$\\
QMAP            &$ 59.0^{+ 6.0}_{- 7.0}$  &$126.0^{+ 54.0}_{- 54.0}$\\
Pyth5/9911419   &$ 22.0^{+ 4.0}_{- 5.0}$  &$ 44.0^{+ 25.0}_{- 15.0}$\\
Pyth5/9911419   &$ 24.0^{+ 6.0}_{- 7.0}$  &$ 75.0^{+ 15.0}_{- 15.0}$\\
Pyth5/9911419   &$ 34.0^{+ 7.0}_{- 9.0}$  &$106.0^{+ 15.0}_{- 15.0}$\\
Pyth5/9911419   &$ 50.0^{+ 9.0}_{-23.0}$  &$137.0^{+ 15.0}_{- 15.0}$\\
Pyth5/9911419   &$ 61.0^{+13.0}_{-17.0}$  &$168.0^{+ 15.0}_{- 15.0}$\\
Pyth5/9911419   &$ 77.0^{+20.0}_{-28.0}$  &$199.0^{+ 15.0}_{- 15.0}$\\
Viper/9910503   &$ 61.6^{+31.1}_{-21.3}$  &$108.0^{+121.0}_{- 78.0}$\\
Viper/9910503   &$ 77.6^{+26.8}_{-19.1}$  &$173.0^{+114.0}_{-101.0}$\\
Viper/9910503   &$ 66.0^{+24.4}_{-17.2}$  &$237.0^{+ 99.0}_{-111.0}$\\
Viper/9910503   &$ 80.4^{+18.0}_{-14.2}$  &$263.0^{+185.0}_{-113.0}$\\
Viper/9910503   &$ 30.6^{+13.6}_{-13.2}$  &$422.0^{+182.0}_{-131.0}$\\
Viper/9910503   &$ 65.8^{+25.7}_{-24.9}$  &$589.0^{+207.0}_{-141.0}$\\
IAC/9907118     &$ 43.0^{+13.0}_{-12.0}$  &$109.0^{+ 19.0}_{- 19.0}$\\
Toco97/9905100  &$ 40.0^{+10.0}_{- 9.0}$  &$ 63.0^{+ 18.0}_{- 18.0}$\\
Toco97/9905100  &$ 45.0^{+ 7.0}_{- 6.0}$  &$ 86.0^{+ 16.0}_{- 22.0}$\\
Toco97/9905100  &$ 70.0^{+ 6.0}_{- 6.0}$  &$114.0^{+ 20.0}_{- 24.0}$\\
Toco97/9905100  &$ 89.0^{+ 7.0}_{- 7.0}$  &$158.0^{+ 22.0}_{- 23.0}$\\
Toco97/9905100  &$ 85.0^{+ 8.0}_{- 8.0}$  &$199.0^{+ 38.0}_{- 29.0}$\\
Toco98/9906421  &$ 55.0^{+18.0}_{-17.0}$  &$128.0^{+ 26.0}_{- 33.0}$\\
Toco98/9906421  &$ 82.0^{+11.0}_{-11.0}$  &$152.0^{+ 26.0}_{- 38.0}$\\
Toco98/9906421  &$ 83.0^{+ 7.0}_{- 8.0}$  &$226.0^{+ 37.0}_{- 56.0}$\\
Toco98/9906421  &$ 70.0^{+10.0}_{-11.0}$  &$306.0^{+ 44.0}_{- 59.0}$\\
MSAM123/9902047 &$ 35.0^{+15.0}_{-11.0}$  &$ 84.0^{+ 46.0}_{- 45.0}$\\
MSAM123/9902047 &$ 49.0^{+10.0}_{- 8.0}$  &$201.0^{+ 82.0}_{- 70.0}$\\
MSAM123/9902047 &$ 47.0^{+ 7.0}_{- 6.0}$  &$407.0^{+ 46.0}_{-123.0}$\\
Boom/9911444    &$ 29.0^{+13.0}_{-11.0}$  &$ 58.0^{+ 17.0}_{- 33.0}$\\
Boom/9911444    &$ 49.0^{+ 9.0}_{- 9.0}$  &$102.0^{+ 23.0}_{- 26.0}$\\
Boom/9911444    &$ 67.0^{+10.0}_{- 9.0}$  &$153.0^{+ 22.0}_{- 27.0}$\\
Boom/9911444    &$ 72.0^{+10.0}_{-10.0}$  &$204.0^{+ 21.0}_{- 28.0}$\\
Boom/9911444    &$ 61.0^{+11.0}_{-12.0}$  &$255.0^{+ 20.0}_{- 29.0}$\\
Boom/9911444    &$ 55.0^{+14.0}_{-15.0}$  &$305.0^{+ 20.0}_{- 29.0}$\\
Boom/9911444    &$ 32.0^{+13.0}_{-22.0}$  &$403.0^{+ 72.0}_{- 77.0}$\\

\hline		
\end{tabular}
}
\end{table}

\begin{figure}[tb] 
\centerline{\epsfxsize=7cm\epsffile{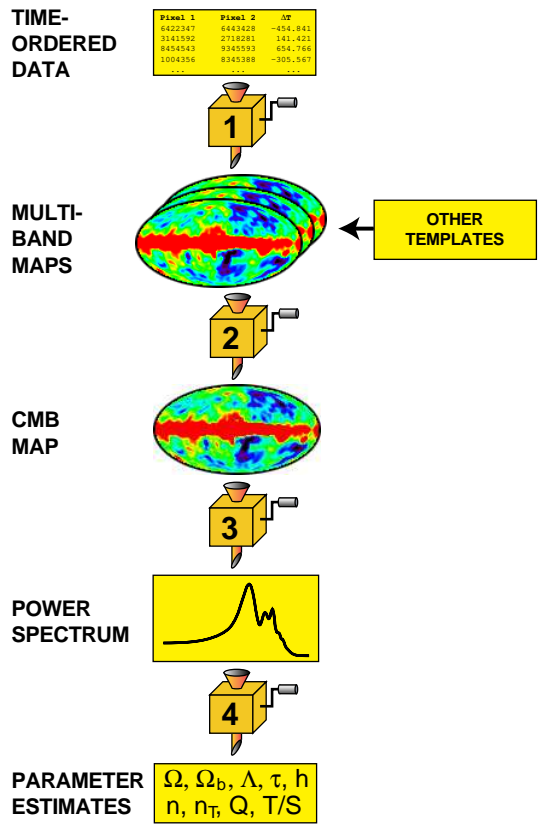}}
\caption{\label{PipelineFig}\footnotesize%
The analysis of a large CMB data set is conveniently 
broken down into four steps: mapmaking, foreground removal, 
power spectrum extraction and parameter estimation.
}
\end{figure}

Comparing these observations with theoretical predictions to achieve this goal
in practice is highly non-trivial, even aside from the experimental challenge
of controlling systematic errors, and is often broken down into several steps,
schematically illustrated in \fig{PipelineFig}:
\begin{enumerate}
\itemsep0cm
\item Compress the time-ordered data set into sky maps at various 
frequencies, so as to minimize the effect of 
correlated detector noise, scan-synchronous offsets, and other non-sky signals
(Wright 1996; Tegmark 1997a).
\item Compress the multi-frequency maps into a single CMB map so as to minimize
the contribution of detector noise and foreground contamination 
(see Tegmark {\etal} 2000 and references therein).
\item Compress this CMB map into measurements of the angular power spectrum
on various angular scales
(Tegmark 1997b; Bond, Jaffe \& Knox 1998), 
a step nicknamed ``radical compression'' by Bond {\etal}
\item Convert these power spectrum measurements into constraints on cosmological 
parameters.
\end{enumerate}
This paper is focused on the last of these four steps,
describing a method and applying it to all currently available data.

Since fast and accurate software is now available for computing how the 
theoretically predicted power spectrum depends on the cosmological parameters,
this last step may at first appear rather trivial: just run a code such as
CMBfast (Seljak \& Zaldarriaga 1996) at a fine grid of points in parameter space and 
perform a $\chi^2$ fit of the corresponding theoretical power spectra to 
the observed data. The problem is that the currently most popular 
cosmological model has of order $N=10$ free parameters, making such 
an $N$-dimensional parameter grid rather huge and unwieldy.
There are also additional challenges related to evaluating the
likelihood function (Bond {\etal} 1998; Bartlett {\etal} 1999) 
that we will discuss in more detail below.

The first analyses based on 
COBE DMR used $N=2$ parameters, the scalar quadrupole normalization 
$\As$ and tilt $\ns$ of the power spectrum
(\eg, Smoot {\etal} 1992; Gorski {\etal} 1994; Bond 1995;  
Bunn \& Sugiyama 1995; Tegmark \& Bunn 1995).
Since then, many dozens of papers have extended this
to incorporate more data and parameters, with recent work including 
Bunn \& White (1997); de Bernardis {\etal} (1997); Ratra {\etal} (1999); Hancock
{\etal} (1998); Lesgourges {\etal} (1999); Bartlett {\etal} (1998); 
Webster {\etal} (1998); Lineweaver \& Barbosa (1998ab); 
White (1998); Bond \& Jaffe (1998); 
Gawiser \& Silk (1998); Contaldi {\etal} (1999),
Griffiths {\etal} 1999; Melchiorri {\etal} (1999);
Rocha (1999).

In an important paper, Lineweaver (1998) 
made the leap up to $N=6$ parameters:
$n_s$, $\As$, the Hubble constant $h$ and
the relative densities $\Oc$, $\Ob$ and $\Ol$
of CDM, baryons and vacuum energy, thereby setting a new standard. 
Tegmark (1999, hereafter T99)
pushed on to $N=8$ by adding the reionization optical depth $\tau$
and the gravity wave amplitude $\At$. 
Efstathiou {\etal} (1999), Efstathiou (1999), 
Bahcall {\etal} (1999),
Dodelson \& Knox (2000) 
and Melchiorri {\etal} (2000) performed analyses with 
different techniques, better data and around 6 parameters, all finding
interesting joint constraints on $\Ol$ and the matter density.
Despite this progress, however, 
a number of issues still need to be improved to do justice to 
the ever-improving data.

Perhaps the most glaring problem is that no closed models 
(White \& Scott 1996) 
have ever been
computed exactly in these analyses, 
except for that of Melchiorri {\etal} (2000), 
since the CMBfast software 
was limited to flat and open models. We remedy this in the present paper
by using version 3.2 of CMBfast (Zaldarriaga \& Seljak 1999), which is
generalized to closed models. A new code by Challinor {\etal} (2000), 
based on CMBfast, also does closed models, agreeing well with CMBfast.
       
Another problem with all previous analyses is that they assumed the 
massive neutrino density $\On$ to be zero, although there is strong 
evidence from both the atmospheric and solar neutrino anomalies that 
$\On>0$. Since these particle physics constraints are only sensitive
to the {\it differences} between the (squared) masses of the various
neutrinos, they do not imply that neutrinos are 
astrophysically uninteresting. 
Indeed, because the CMB and matter power spectra 
can place some of the most stringent 
upper limits on neutrino masses (Hu {\etal} 1998), 
it would be a real pity 
to omit this aspect of the analysis.
Just as increasing $\Oc$
suppresses the acoustic peaks, 
increasing $\On$ suppresses does so by a comparable amount.
Indeed, these two parameters become nearly degenerate for large $\On$, 
corresponding to neutrinos massive enough to be fairly nonrelativistic at 
the relevant redshifts, so the inclusion of neutrinos will, among other things, 
weaken the lower limit on $\Oc$.

Another weakness of the T99 analysis was that it assumed 
that the relative amplitude 
$r\equiv\At/\As$ 
of gravity waves was linked to the tensor spectral index by the
inflationary consistency relation 
$r = -7\nt$ (Liddle \& Lyth 1992),
although one of the most exciting applications of CMB data 
will be to test this relation. We will remedy both of these problems 
by extending our parameter space to $N=10$ dimensions, including 
both $\On$ and $\nt$ as free parameters.

Finally, as we will discuss at length below, there are 
a number of areas where accuracy has been unsatisfactory and 
can be substantially improved.

The rest of this paper is organized as follows.
We describe our method in \sec{MethodSec}, apply it to the available data in
\sec{ResultsSec} and summarize our conclusions in \sec{ConclusionsSec}.
Some technical details regarding marginalization are derived in the Appendix.

\section{METHOD}

\label{MethodSec}\

\begin{figure}[tb] 
\vskip-1.0cm
\centerline{\epsfxsize=9.5cm\epsffile{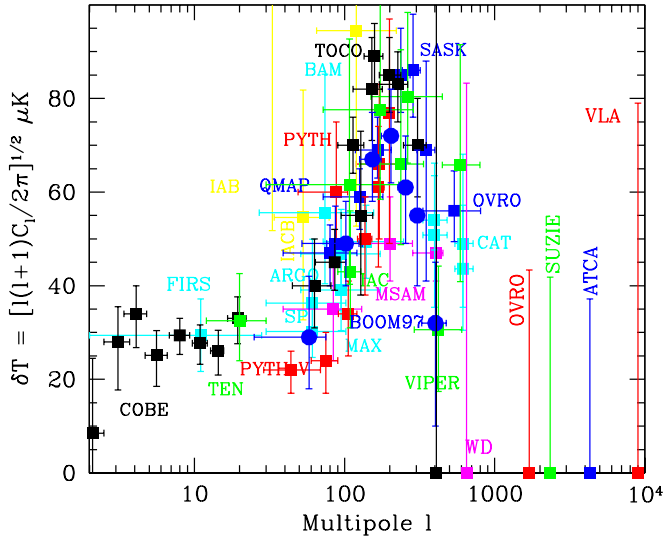}}
\vskip-1.0cm
\caption{\label{DataFig}\footnotesize%
The band power measurements used.
}
\end{figure}

\subsection{The problem}

Our data consists of the $n=65$ band power measurements $\dT^2_i$
listed in Table 1 and shown in \fig{DataFig}, $i=1,...,n$.
The band power measurement $d_i\equiv \dT^2_i$ probes a weighted average
of  $\delta T_\l^2\equiv\l(\l+1)C_\l/2\pi$,
\beq{WindowEq}
\expec{d_i} = \expec{\dT^2_i} =
\sum_\l {1\over\l}W^i_\l \delta T_\l^2,
\eeq
where $W^i_\l$ is the band-power window function (distinct from
the variance window function; see Knox 1999).
These known weights $W^i_\l$
reflect which angular scales the measurement is sensitive to.

The power spectrum in turn depends on our vector of cosmological 
parameters $\p$ in a complicated fashion $C_\l(\p)$ that we use CMBfast to
compute. The scatter in the relation between $d_i$ and $\expec{d_i}$
due to detector noise and sample variance is described by 
a likelihood function $\L_i(d_i;C_\l(\p))$, the probability
distribution for $d_i$ given $\p$.
If the errors in the different data points were all independent, 
then the combined likelihood of observing the set of all data
given $\p$ would be simply
\beq{Leq}
\L(\data;\p) = \prod_{i=1}^n \L_i(d_i;C_\l(\p)).
\eeq
This is complicated by the fact that some measurements are correlated, 
as will be discussed in \sec{LikelihoodSec}.

Our problem is to evaluate this likelihood function in the 10-dimensional 
parameter space that $\p$ inhabits. To obtain Bayesian constraints on 
individual parameters or joint constraints on interesting
pairs (such as $\Om$ and $\Ol$), we then marginalize over the remaining
parameters with appropriate priors.

\subsection{Breaking it into four sub-problems}

If we had infinite computing resources, the solution would be 
straightforward: compute the theoretical CMB power spectrum $C_\l(\p)$
with the CMBfast software (Seljak \& Zaldarriaga 1996) and the corresponding
likelihood at a fine grid of points in 
the $N$-dimensional parameter space.  
In practice, this is inconvenient. With $M$ grid points in each 
dimension, $M^N$ power spectra must be computed.
Even if we take $M$ as low as 10, the amount of 
work thus grows by an order of magnitude for 
each additional parameter. With 1 minute per power spectrum
calculation, $N=10$ would translate to over $10^4$ years of 
CPU time. 

Fortunately, the underlying physics
(see {\eg} Hu {\etal} 1997 for a review) 
allows several numerical simplifications to be made.
We will adopt the approximation scheme used in T99 with additional
improvements as described below.
Our method conveniently separates into four separate steps.
\begin{itemize}
\item {\bf Step 1:} Run CMBfast many times for three particular subsets of
the parameter grid. The results are three large files: one with tensor 
power spectra, one with scalar power spectra for $\ell\simlt 100$
and one with scalar power spectra for $\ell\simgt 100$.
\item {\bf Step 2:} Interpolate these spectra 
onto larger subsets of the
parameter grid. The results are two huge files with 7-dimensional
model grids, one for scalars and one for tensors.
These two files allow any power spectrum in the full 10-dimensional 
model grid to be computed almost instantaneously. 
\item {\bf Step 3:} Compute and save the likelihood $\L$ for each model.
\item {\bf Step 4:} Perform 10-dimensional interpolation and marginalize
to obtain constraints on individual parameters, constraints in the
$(\Om,\Ol)$-plane, \etc
\end{itemize}
Below we will describe each of these four steps in turn.
Before doing this, however, it is interesting to contrast this 
``huge grid'' approach with an alternative strategy.
Dodelson \& Knox (2000) and Melchiorri {\etal} (2000)
performed their analyses without computing and 
storing such a grid. Instead, they found the maximum-likelihood 
parameter vector $\p$ by a direct numerical maximum search, 
computing power spectra with CMBfast on the fly as needed. 
Similarly, constraints in say the $(\Om,\Ol)$-plane were obtained by
performing a numerical maximum search over the remaining parameters
for each $(\Om,\Ol)$ grid point.
One drawback of this approach is that everything needs to be repeated 
from scratch if the data set is changed, whereas steps 1 and 2 
in our method are independent of the data set and need only be done
once and for all. The same drawback applies to exploring different
priors.
There is also no guarantee that CMBfast gets run fewer times with this
direct search approach, as a numerical search in the high-dimensional space
tends to require large numbers of likelihood evaluations
(Dodelson 1999, private communication; see also Hannestad 1999).

\subsection{Parameter space}

We choose our 10-dimensional parameter vector to be
\beq{pEq}
\p\equiv(\tau,\Ok,\Ol,\oc,\ob,\on,\ns,\nt,\As,\At),
\eeq
where the physical densities 
$\omega_i\equiv h^2\Omega_i$, $i=$cdm, b, $\nu$. The advantage of this 
parameterization (see Bond {\etal} 1997)
will become clear in \S\ref{HighLowSec}.
$\Ok$ is the spatial curvature, so 
in terms of these parameters,
\beq{hEq}
h = \sqrt{\oc+\ob+\on\over 1-\Ok-\Ol}.
\eeq
This parameter space is identical to that used in T99 except that we have 
added $\on$ and replaced $h$ by $\Ol$ as a free parameter.

We wish to probe a large enough region of parameter space to cover
even quite unconventional models. This way, constraints from non-CMB
observations can be optionally included by explicitly multiplying 
$\L(\p)$ by a Bayesian prior after Step 3 rather than being 
hard-wired in from the outset.
To avoid prohibitively large $M$,
we use a roughly logarithmic
grid spacing for $\om$, $\ob$ and $\on$, 
a linear grid spacing for $\Ok$ and $\Ol$,
a hybrid for $\tau$, $\on$, $\ns$ and $\nt$, 
and (as described below) no grid at all for $\As$ and $\At$.
We let the parameters take on the following values:
\begin{itemize}
\item $\tau=0, 0.05, 0.1, 0.2, 0.3, 0.5, 0.8$ 
\item $\Ol=-1.0, -0.8, -0.6, -0.4, ...., 1.0$ 
\item $\Ok$ such that $\Om\equiv 1-\Ok-\Ol = 0.2, 0.4, ..., 2.0$
\item $\oc=0.02, 0.03, 0.05, 0.08, 0.13, 0.2, 0.3, 0.5, 0.8$ 
\item $\ob=0.003, 0.005, 0.008, 0.013, 0.02, 0.03, 0.05, 0.08, 0.13$ 
\item $\on=0, 0.02, 0.05, 0.08, 0.13, 0.2, 0.3, 0.5, 0.8$ 
\item $\ns=0.50, 0.70, 0.90, 1.00, 1.10, 1.20, 1.30, 1.50, 1.70$ 
\item $\nt=-1.00, -0.70, -0.40, -0.20, -0.10, 0$ 
\item $\As$ is not discretized
\item $\At$ is not discretized
\end{itemize}
Note that the extent of the $\Ok$-grid depends on $\Ol$, giving 
a total of $10\times 11=110$ points in the $(\Om,\Ol)$-plane.
Our discrete grid thus contains 
$7\times 110\times 9\times 9\times 9\times 9\times 6=30,311,820$ models.
As will become clear from our discussion below, the main limitation on
this grid size is the disk space used in Step 2 rather than the 
CPU time used in Step 1, so it will probably be desirable to further
refine it as CMB data gets better.

\subsection{Separating scalars and tensors}

If we were to run CMBfast in the standard way, computing scalar and tensor
fluctuations simultaneously, we would have to explore a 9-dimensional 
model grid since only $\As$ drops out as an overall normalization factor. 
Instead, we compute 
the scalar fluctuations $\Cl^{scalar}$ and 
the tensor fluctuations $\Cl^{tensor}$
separately, normalize them to both have a quadrupole of unity, and 
compute the combined power spectrum as
\beq{TensorComboEq}
\Cl = \As\Cl^{scalar} + \At\Cl^{tensor}.
\eeq
We therefore only need to compute two 
7-dimensional grids with CMBfast, one over
$(\tau,\Ok,\Ol,\oc,\ob,\on,\ns)$
and the other over
$(\tau,\Ok,\Ol,\oc,\ob,\on,\nt)$. The other advantage of calculating 
scalars and tensors separately is that tensors only need to be
calculated up to an $l$ of 400, which saves additional time. 

Allowing 1 minute per model, the scalar grid alone would still
take about 10 years of CPU time. Most models take substantially longer
to run, since reionization, curvature and neutrinos slow CMBfast down.
It is therefore useful to take advantage of the underlying physics to make further
simplifications.

\subsection{Separating small and large scales}
\label{HighLowSec}


The tensor power spectrum depends only 
weakly on $\oc$, $\ob$ and $\on$.
We therefore compute 
the tensor power spectrum
with the fine grid restricted to 
$(\tau,\Ok,\Ol,\nt)$, 
using only ultra-course three-point grids for 
$\oc$, $\ob$ and $\on$. 
We then fill in the rest of the 
$(\oc,\ob,\on)$-values using cubic spline interpolation.

The scalar power spectrum $\Cl$ for $\l\ll 100/\Om^{1/2}$ corresponds
to fluctuations on scales outside the
horizon at recombination. This makes it almost independent of 
the causal microphysics that create the familiar acoustic peaks, 
\ie, independent of $\om$, $\on$ and $\ob$. We therefore compute 
the scalar power spectrum
on large scales with the fine grid restricted to 
$(\tau,\Ok,\Ol,\ns)$, using only ultra-course 
three-point grids for $\oc$, $\ob$ and $\on$ to 
to pick up weak residual effects aliased down 
from larger $\l$. 
We then fill in the rest of the $(\oc,\ob,\on)$-values
using cubic spline interpolation.

For the remaining (high $\l$) part of the power spectrum, 
more radical simplifications can be made.
First of all, the effect of reionization is mainly
an overall suppression of $\Cl$ by a constant factor 
$e^{-2\tau}$ on these small scales.
Second, the effect of changing both $\Ok$ and $\Ol$ 
is merely to shift the power spectrum sideways. This is because the
acoustic oscillations at $z\simgt 1000$ 
(at which time $\Ok\approx\Ol\approx 0$ regardless of their present value)
depend only on 
$\om$, $\ob$ and $\on$, and the geometric projection 
of these fixed length scales onto angular scales $\theta$ in the sky
obeys 
$\theta\propto 1/\dA$, where
$\dA$ is the angular diameter distance to the last scattering surface.
In T99 and Efstathiou {\etal} (1999), 
$\dA$ was estimated analytically by integrating out to the
redshift of last scattering given by the fit of Hu \& Sugiyama (1996).
Since CMBfast automatically computes this quantity anyway, we 
eliminate this approximation by simply using this numerical value.

$\Om$ and $\Ol$ also modify the
late integrated Sachs-Wolfe effect, but this is important
only for $\l\simlt 30$ (Eisenstein {\etal} 1999). 
The only other effect is a small correction due to gravitational lensing
(Metcalf \& Silk 1997; Stompor \& Efstathiou 1999), which we ignore here because of  
the large error bars on current small-scale data.
To map the model $\p^*$ into the model $\p$ with all parameters 
except $\tau$, $\Ok$ and $\Ol$ unchanged,
we thus multiply its high $\l$ power spectrum
by $e^{2(\tau^*-\tau)}$ and shift it to the right by an $\l$-factor of 
$\dA/\dA^*$.
 
We therefore adopt the following procedure for the first two steps. 
In Step 1, we compute
\begin{itemize}
\item scalar power spectra out to $\l=5000$ for the subgrid with
$\tau=\Ok=\Ol=0$ (merely 6,561 models),
\item scalar power spectra out to $\l=400$
with the subgrid restricted to
$\tau=0, 0.1, 0.8$, 
$\om=0.02, 0.2, 0.8$, 
$\ob=0.003, 0.02, 0.13$,
$\on=0, 0.2, 0.8$
(80,190 models), and
\item tensor power spectra with the matter densities restricted to 
this same subgrid (80,190 models).
\end{itemize}

In Step 2, we use cubic spline interpolation separately for each $\l$
to extend the tensor models and the low-$\ell$ scalar models 
to the full parameter grid. 
To account for the effects of $\tau$, $\Ok$ and $\Ol$, 
we then shift the high-$\ell$ scalar models vertically and horizontally 
as described above and splice them together with the corresponding
low-$\l$ models at a cutoff value $\lstar$.
For a given model, we choose $\lstar$ to be 100 multiplied by the horizontal 
shifting factor. In other words, the high-$\ell$ model always gets spliced
at the location that corresponded to $\l=100$ before shifting it sideways,
so open models get spliced at higher $\l$ and closed at lower.
When computing the low-$\ell$ models in Step 1, we therefore 
adjust the accuracy flag ``ketamax'' in CMBfast 
to be 400 times this same shifting factor.

The public releases of CMBfast normalize the power spectra $C_\l$ 
to COBE automatically. 
This normalization scheme is not appropriate for our merging technique,
since we need a convention
independent of the cosmological parameters so that when we combine the
high and low grids, 
the relative normalization of the models is correct. To achieve this, we
removed the COBE normalization from CMBfast and normalized the
power spectrum in both the flat and non-flat codes to agree on scales 
much smaller than the curvature scale.  

For the reader interested in implementing this scheme, it is worth noting
that almost all the time in Step 1 is spent on the low scalar grid.
For this grid, substantial time is saved by only computing the 
power spectrum for the low $\l$-values where it is needed.
Note that the loop over tilts ($\ns$ or $\nt$) is essentially
free, since CMBfast can compute multiple tilts simultaneously.
The only reason we have used so few tilt values is because 
of disk space considerations in Steps 2 through 4.
Including various test runs, we filled up more than
half of a 200 GB disk array.

\subsection{Testing step 2}

To test the accuracy of the resulting scalar and tensor model grids
produced in Step 2, we drew a random sample of $\sim 10^3$ of 
the models and recomputed them from scratch with CMBfast.
For most models, we found our results to be accurate to a within 
a few percent. The remainder generally had very early reionization
(high $\tau$ and low $h\Ob$),
which causes a broad bump of regenerated power from motions on the
new last scattering surface. Since our approximation simply suppresses the
small scale power by $e^{-2\tau}$, it therefore underpredicts the
power on the angular scale corresponding to the horizon size at reionization.
In addition, the interpolation performed poorly at the lowest $\l$
for some quite crazy models, which could be remedied by running CMBfast 
on a finer grid.

As data quality improves further, it will probably be worthwhile to 
simply include $\tau$ explicitly in the high-$\l$ grid.
In this case, the remaining errors introduced by our
approximation scheme can of course be continuously reduced to zero by
refining the $(\oc,\ob,\on)$-grid for low 
$\l$ and shifting the splicing point upwards from $\l\sim 100$.

\subsection{Step 3: computing likelihoods}
\label{LikelihoodSec}

We use the CMB data and window functions listed in Table 1
and shown in \fig{DataFig}. 
This is taken from the compilation of Lineweaver (1998) with the
addition of the new results from QMAP
(Devlin {\etal} 1998; Herbig {\etal} 1998; de Oliveira-Costa {\etal} 1998),
MSAM (Wilson {\etal} 1999), 
Toco (Torbet {\etal} 1999; Miller {\etal} 1999), 
Python V (Coble 1999), Viper (Peterson {\etal} 2000) and
Boomerang (Mauskopf {\etal} 1999). 
For an up-to-date annotated compilation of all current data, 
see Gawiser \& Silk (2000).
For the COBE data, we use the exact window function from Tegmark (1997b).
In all other cases, we approximate the window functions by a Gaussian
of FWHM=$\lhi-\llo$ from Table 1.
This approximation does not appear to have much of an effect on the
results: we repeated the analysis with the much more extreme approximation
where the windows are delta functions at $(\llo+\lhi)/2$ and obtained
essentally unchanged results.
Knox \& Page (2000) compared full window functions with delta functions
and came to the same
conclusion.

\def\Cuc{\C^{\rm (meas)}}
\def\Cic{\C^{\rm (ical)}}
\def\Csc{\C^{\rm (scal)}}

As discussed in great detail by Bond, Jaffe \& Knox (1998) and also by 
Bartlett {\etal} (1999), 
an accurate calculation of the likelihood function 
$\L(\data|\p)$ is nontrivial. If the band-power measurement
$d_i$ is a quadratic function of the sky temperatures measured by the
experiment in question, then $\L_i(d_i;C_\l(\p))$ is a generalized 
$\chi^2$ distribution when viewed as a function of $d_i$ 
(Wandelt {\etal} 1998), but sufficient details to compute this function
exactly are rarely published when band power measurements are released.
Useful approximations have therefore been derived that require only 
the asymmetry between upper and lower error bars as input 
(Bond, Jaffe \& Knox (1998), Bartlett {\etal} (1999).
The former approximation is implemented by a nice publicly available
package called RADPACK, maintained by Lloyd Knox at 
{\it http://flight.uchicago.edu/knox/radpack.html}, which was used in 
the analyses of Dodelson \& Knox (2000) and Melchiorri {\etal} (2000).
In this paper, we will stick with the cruder Gaussian approximation
\beq{chi2eq}
\L(\T;C_\l(\p))\approx
e^{-{1\over 2}(\T-\expec{\T})^t\C^{-1}(\T-\expec{\T})},
\eeq
where $\T$ is the vector of measurements $d_1,d_2,...d_n$ 
and $\C$ is the associated $n\times n$ covariance matrix of
measurement errors.
This means that the full likelihood function
$\L=e^{-\chi^2/2}$, where $\chi^2$ is simply the chi-squared goodness 
of fit of the model to the data.

We have chosen to keep things 
this simple because we are currently unable to eliminate a 
third major source of inaccuracy: many of the recent multi-band measurements
released (which dominate the constraining power) have non-negligible correlations
between their different bands, but these correlations have not yet been
published by the experimental teams. An alternative approach would be to
convert these data sets to uncorrelated measurements, as was done with 
the 8 COBE points we use.
In the interim, an alternative
is to simply use only 
those experiments which either have very small correlations,
or significant correlations which are publically available,
as was done in Dodelson \& Knox (2000) and Knox \& Page (2000).

We model $\C$ as a sum of three terms, 
$\C=\Cuc + \Csc + \Cic$, corresponding to
measurement errors, 
source calibration errors and 
instrument calibration errors, 
respectively.
$\Cuc$ reflects the part of the errors which are uncorrelated
between the different experiments and is due to detector noise and 
sample variance. We approximate it by
\beq{uncorrCueq}
\Cuc_{ij}\equiv\delta_{ij}\sigma_i^2,
\eeq
where $\sigma_i$ is defined as the average of the upper and lower error
bars quoted for $d_i\equiv\dT^2$ (not for $\dT$) in Table 1. 

The last two terms reflect the correlations between measurements due
to calibration errors. $\Cic$ is the part specific to a single 
multi-band experiment and $\Csc$ is the part that is correlated with
other experiments that are calibrated off of the same (slightly uncertain) source.
Both QMAP and Saskatoon calibrate off of Cass A, 
and we assume that a 8.7\% error due to the flux uncertainty
of this object is common to these experiments.
MAT, MSAM and Boomerang all calibrate off of Jupiter.
To be conservative, we assume that the full 5\% 
calibration uncertainty from 
Jupiter's antenna temperature is shared by these experiment.
The true correlation should be lower, since the three experiments 
observed Jupiter at different frequencies.
The remaining multi-band experiments do not have any such 
inter-experiment correlations:
COBE/DMR calibrated off of the dipole, Viper off of the moon
and Python V off of internal loads.
This contribution to the noise matrix is therefore
\beq{CscEq}
\Csc_{ij}\equiv (2s_{ij})^2 d_i d_j,
\eeq
where 
\beq{rEq}
s_{ij} = \cases{
8.7\% &if $i$ and $j$ refer to QMAP or Saskatoon,\cr
5\%   &if $i$ and $j$ refer to MAT, MSAM or Boom,\cr
0     &otherwise.\cr
}
\eeq
The factor of 2 in \eq{CscEq} stems from the fact that the percentage 
error on $\dT_i^2$ is roughly twice that for $\dT_i$ as long as 
it is small. 
Similarly, the remaining term is 
\beq{uncorrCieq}
\Cic_{ij}\equiv (2r_{ij})^2 d_i d_j,
\eeq
where $r_{ij}=0$ if $i$ and $j$ refer to different experiments.
If band powers $i$ and $j$ are from the same experiment, then
$r_{ij}$ is the quoted quoted calibration error with
the source contribution $s_{ij}$ subtracted off in quadrature.
We use 
$r=$0.063 for Saskatoon, 
7.9\% for QMAP, 
14\% for Python V,
8\% for Viper, 
8.7\% for Toco 97,
6.2\% for Toco 98,
0 for MSAM
and
6.4\% for Boomerang.

There is certainly ample room for improvement of in this 3rd step.
To put all these statistical issues in perspective, however, 
the authors feels that an even more pressing challenge will be to test the 
data sets for systematic errors, \eg, by comparing them pairwise
where they overlap in sky coverage and angular resolution 
(Knox {\etal} 1998; Tegmark 1999a).

\subsection{Step 4: Marginalizing}

For a Bayesian analysis, the 10-dimensional likelihood should be multiplied
by a prior probability distribution reflecting all non-CMB information,
then rescaled so that it integrates to unity and can be interpreted as
a probability distribution.
To obtain constraints on some subset of the parameters 
($\Ok$ and $\Ol$, say), one would then marginalize over all other
parameters by integrating over them.
Such a direct integration was performed by Efstathiou {\etal} (1999)
where the parameter space had fewer dimensions. Since such integration
is quite time-consuming in a high-dimensional space, most other 
multi-parameter analyses published have adopted the alternative approach
of maximizing rather than integrating over the unwanted parameters. 
For instance, the reduced likelihood function for $\tau$ is obtained
by looping over a grid of $\tau$-values and choosing the remaining parameters
so that they maximize the likelihood in each case.
These two approaches are equivalent if the full likelihood function is 
a multivariate Gaussian, as shown in Appendix A. 
If Gaussianity is a poor approximation, 
the maximization approach can tend to underestimate the error bars
(Efstathiou {\etal} 1999). The Gaussianity approximation is
indeed a poor one at the moment, especially for the case with no 
priors, but it should gradually improve as
future data and non-CMB priors reduce the size of the allowed parameter 
region.

\begin{figure}[tb] 
\vskip-1.2cm
\centerline{\epsfxsize=9.5cm\epsffile{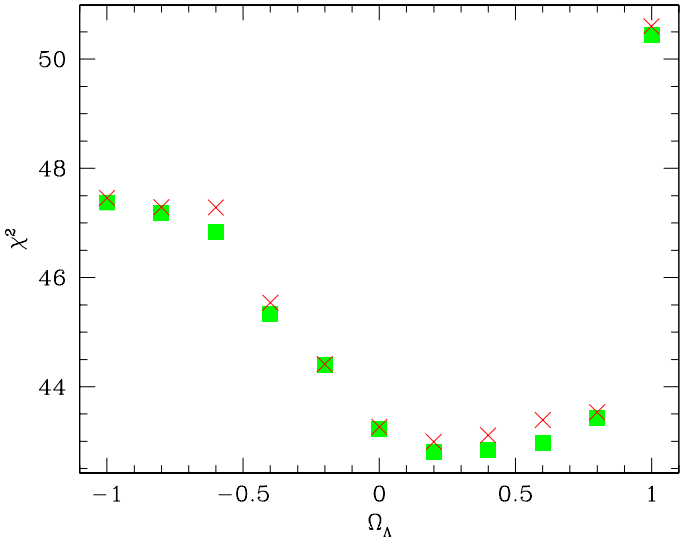}}
\vskip-1.0cm
\caption{\label{MargMethodFig}\footnotesize%
Marginalization method comparison.
$\chi^2$ is plotted as a function of $\Ol$ when maximizing 
over all other parameters with no priors. The squares show the
result of using multidimensional spline interpolation when maximizing
and the crosses show the result of simply picking the smallest $\chi^2$-value
in the model grid. Note that a seemingly small error of unity 
in $\chi^2$ changes the likelihood by a factor of 1.6.
}
\end{figure}

In the published grid-based implementations of the maximization method
(\eg, Lineweaver 1998; T99),
the minimization was performed by simply looking at the
likelihoods in the pre-computed model grid and picking the largest 
one. Since the true maximum does generally not reside exactly at a grid point, 
this method always underestimates the true maximum. Unfortunately, 
the magnitude of this underestimation will vary in a rather random way,
depending on how close to the constrained maximum happens to be to
the nearest grid point. This effect can cause jagged-looking and
somewhat misleading results,
as shown in \fig{MargMethodFig}. Note that even an error as small as
0.5 in $\chi^2$ changes the likelihood by more than 20\%.
Some of the jaggedness/ringing seen in the plots in, \eg, 
Lineweaver (1998) and T99 is likely to be due to this effect.
In contrast, the ringing seen in many of the 
$(\Om,\Ol)$ exclusion plots
further on in this paper is a purely 
cosmetic problem, due to instability in the IDL interpolation
routine used to generate the contour plots.

The problem at hand is to find the maximum of some hypersurface
in a high-dimensional space. It is easy to see that if we approximate
the surface by multilinear interpolation between the grid points
where we know its height, we will recover this unsatisfactory method, 
since the interpolated surface can only have maxima at grid points.
We have chosen to use cubic spline interpolation instead. 
As seen in 
\fig{MargMethodFig}, this works substantially better and eliminates 
the random jaggedness of the simpler method.

For the reader interested in implementing this method, we give some 
additional practical details below. 
Other readers may wish to skip directly to the next subsection.

We perform the cubic spline interpolation and subsequent 
maximization one dimension at a time. Just as for multilinear 
interpolation, the result of this procedure is independent of the order
in which we interpolate over the different parameters.
We start by maximizing over the scalar and tensor normalizations, 
which is readily done analytically since $\chi^2$ depends 
quadratically on $\As$ and $\At$.
We save the remaining 8-dimensional grid in a huge file together
with the optimal values of $\As$ and $\At$ and the corresponding 
$\chi^2$ value.
To marginalize over any given parameter $p_i$, we first sort
this file so that this parameter varies fastest. 
In each block where the remaining parameters are fixed, 
we then spline over this parameter and find the maximum $p_i^*$
analytically from the spline coefficients.
Since it is interesting to keep track of the physical 
parameters of the best fit models, we save not only the
$\chi^2$-value but also the other parameter values 
spline interpolated to the point where $p_i=p_i^*$, 
replacing the entire block of models in the file 
by this interpolated one.

We found that when $\chi^2$ varies rapidly, a standard cubic
spline occasionally causes unwanted oscillations. Such a rapid 
rise in $\chi^2$ occurs only in the extreme parts of the parameter
grid that we do not care about (since they are completely ruled out),
yet the resulting ringing easily propagates to the region that we
are interested in near the minimum. We therefore adopted 
a scheme where we through away irrelevant distant points before splining
if they were too extreme. Specifically, before performing a 1-dimensional 
cubic spline, we first located the lowest grid point.
We then included all points to the left of it until we reached one
whose $\chi^2$ was higher by 10 or more. 
Points to the right were included analogously.
We found this simple scheme to work quite well in practice. Indeed, the
slight wiggliness of the contour plots shown in the next section is
caused mainly by the plotting software itself (the 2D interpolation
routine of IDL), not by our marginalization from 10 to 2 dimensions.

\section{RESULTS}

\label{ResultsSec}

\subsection{Basic results}

To avoid having our constraints severely diluted by 
``silly'' models, we include two prior pieces of information
when presenting our basic results. We assume that the Hubble parameter
$h=0.65\pm 0.07$ at $1-\sigma$ (see Freedman 1999 for a recent 
review of $h$-measurements) and
that the baryon density 
$\ob=h^2\Ob\approx 0.02$ 
(Burles {\etal} 1999
report $\ob=0.019\pm 0.0024$, and we approximate the $\ob$ error bars by zero
since they are much smaller than our $\ob$ grid spacing).
This value of $\ob$ is roughly consistent with
that measured by Wadsley {\etal} (1999) using the Helium Lyman-Alpha Forest.
We assume that the error distribution for $h$ is Gaussian.

\begin{figure}[tb] 
\vskip-1.0cm
\centerline{\epsfxsize=9.5cm\epsffile{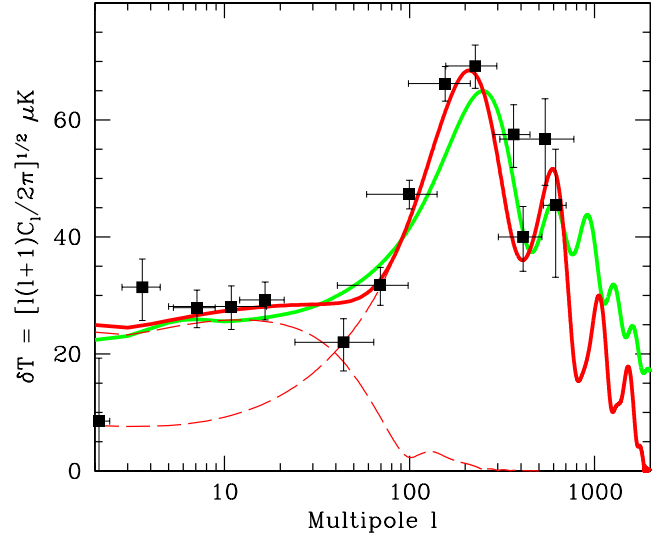}}
\vskip-1.0cm
\caption{\label{BestFitFig}\footnotesize%
The best fit model is shown for the case of 
no prior (solid red/dark grey)
and with the priors $h=0.65\pm 0.07$, $h^2\Ob=0.02$
and $\tau=r=0$ (solid green/light grey).
The dotted lines show the decomposition of the former curve
into scalar and tensor fluctuations. The model parameters are
listed in Table 2.
Although all 65 measurements were used in the fits, 
they have been averaged into 14 bands in this plot to 
avoid cluttering. The band powers whose central $\l$-value 
fell into any given band were average with minimum-variance weighting,
and their corresponding window functions were averaged as well.
This binning was used only in this plot, not in our analysis.
}
\end{figure}

\begin{table}
\def\Qrmsps{Q_{rms,ps}}
\def\lm#1#2#3{$#1^{+#2}_{-#3}$}
\def\na{$-$}
\def\lpeak{\l_{peak}}

\bigskip
\noindent
{\footnotesize
{\bf Table 2} -- Maximum-likelihood values and 95\% confidence limits
}
\smallskip
{
\begin{tabular}{|l|ccc|ccc|}
\hline
			&\multicolumn{3}{c|}{10 free parameters}
			&\multicolumn{3}{c|}{$h$ \& $\ob$ prior}\\
Quantity		&Min	&Best	&Max	&Min	&Best	&Max\\
\hline
$\tau$			&0.0	&0.0	&$-$	&0.0	&0.0	&$-$\\	
$\Ok$			&$-$1.74&$-$1.03&0.49	&$-$0.24&.09	&0.38\\	
$\Ol$			&$-$	&.16	&$-$	&$-$0.19&.67	&0.89\\	
$h^2\Oc$		&0.0	&.53	&$-$	&0.0	&.036	&0.30\\	
$h^2\Ob$		&.11	&.13	&$-$	&{\it .02}&{\it .02}&{\it .02}\\	
$h^2\On$		&0.0	&.012	&$-$	&0.0	&.051	&.29\\	
$\ns$			&.55	&1.69	&$-$	&0.80	&1.05	&1.53\\	
$\nt$			&$-$	&0.00	&$-$	&$-$	&0.03	&$-$\\	
\hline		
\end{tabular}
}
\end{table}

%

The parameters of the best fit model
are listed in Table 2 both with and without these priors.
The corresponding no-prior power
spectrum is shown in \fig{BestFitFig} together
with the ``vanilla'' version with 
the above-mentioned priors and $\tau=r=0$.
As can be seen, 
the fitting procedure uses the additional freedom 
to match features in the data in quite amusing ways.
Since the data dip at $\l\sim 50$ and rise very sharply
thereafter, a feature that simpler models cannot match,
the minimization procedure finds the best fit model
to have a dramatic blue-tilt ($\ns\sim 1.7$) and 
almost the entire COBE signal due to gravity waves.
Although this particular model is ruled out by other constraints
--- for instance, 
primordial black hole abundance (Green {\etal} 1997) and 
spectral distortions (Hu {\etal} 1994) give upper bounds $\ns\simlt 1.3$) ---
it illustrates the importance of fitting for all 10 parameters 
jointly. Indeed, it is the inclusion of gravity waves in
our models that makes the constraints on $\ns$ so weak.

\begin{figure}[tb] 
\centerline{\epsfxsize=9.0cm\epsffile{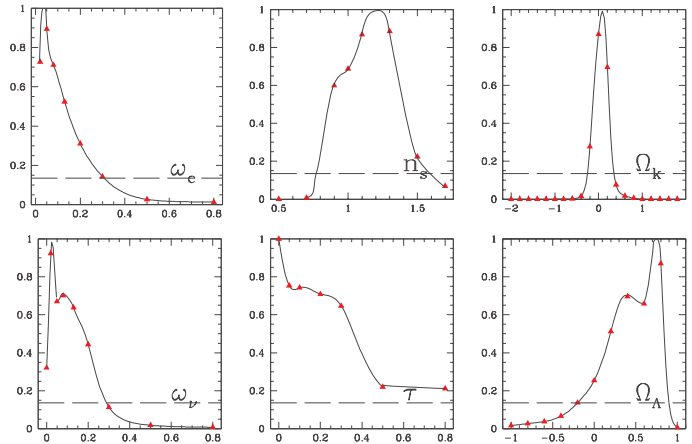}}
\caption{\label{1Dfig}\footnotesize%
The marginalized likelihood is shown for six individual parameters
using all 65 band power measurements and priors only
from nucleosynthesis ($h^2\Ob=0.02$)
and the Hubble parameter ($h=0.65\pm 0.07$).
The $2\sigma$ limits (see Table 2) are roughly where 
the curves cross the horizontal lines.
}
\end{figure}

The 1-dimensional likelihood functions for six of the best constrained 
parameters are shown in \fig{1Dfig}, marginalized over the 
other 9 parameters. 
Although none of these parameters are very tightly constrained, 
it is encouraging that CMB observations are already sufficiently
powerful to place upper and lower limits on 
$\Om$, $\Ol$ and $\ns$ at the $2-\sigma$ level.
Because $\oc$ and $\on$ are by definition non-negative, these 
density parameters are also bounded from both sides. 
On the other hand, better data will be required 
to place interesting constraints on $\tau$,
since this parameter is almost degenerate with the overall normalization
(see, \eg, Eisenstein {\etal} 1999). 
The best constrained parameter so far is seen to be 
the spatial curvature $\Ok$, with $-0.24<\Ok<0.38$ at 95\% confidence.
For comparison, using Figure 2 in Dodelson \& Knox (2000) to read 
off the point where  the likelihood drops to $e^{-2^2/2}\approx 0.14$ 
gives the 95\% upper limit $\Ok<0.38$.
Although the exact numerical agreement is likely to be 
coincidental (since we use more data, etc),
this is nonethetheless very reassuring evidence that the basic
result is robust.
       
\begin{figure}[tb] 
\centerline{\epsfxsize=9.0cm\epsffile{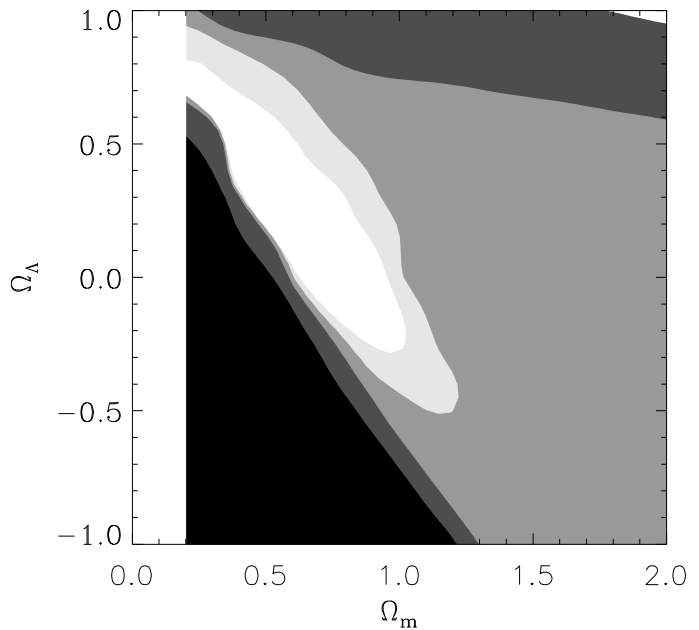}}
\caption{\label{all65_2Dfig}\footnotesize%
The regions in the $(\Om,\Ol)$-plane 
that are ruled out at $2\sigma$ using all the data are shown 
using no priors (red/dark grey), 
the prior $h=0.65\pm 0.07$ (orange red/grey),
the additional nucleosynthesis constraint $h^2\Ob=0.02$ (orange/light grey)
and the additional constraints $r=\tau=0$ (yellow/very light grey). 
}
\end{figure}

\begin{figure}[tb] 
\centerline{\epsfxsize=9.0cm\epsffile{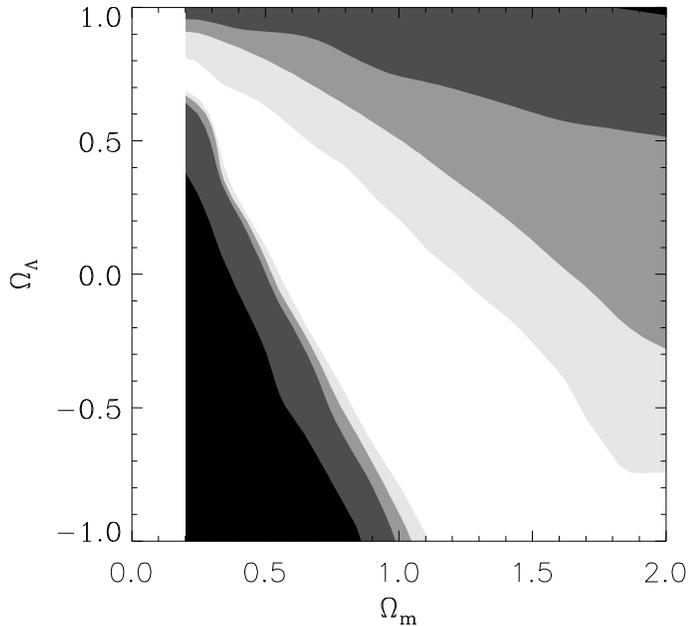}}
\caption{\label{lyman2Dfig}\footnotesize%
Same as previous figure, but
using only the COBE and the ``East Coast''
data sets (Saskatoon, QMAP and Toco).
}
\end{figure}

\begin{figure}[tb] 
\centerline{\epsfxsize=9.0cm\epsffile{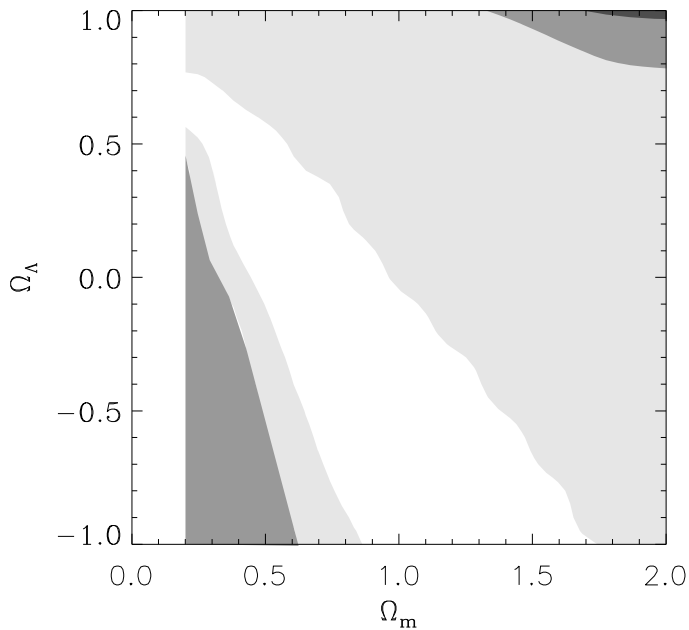}}
\caption{\label{snake2Dfig}\footnotesize%
Same as Figure~\ref{all65_2Dfig}, but
using only COBE and ``Snake'' data sets
(Python V and Viper).
}
\end{figure}

\begin{figure}[tb] 
\centerline{\epsfxsize=9.0cm\epsffile{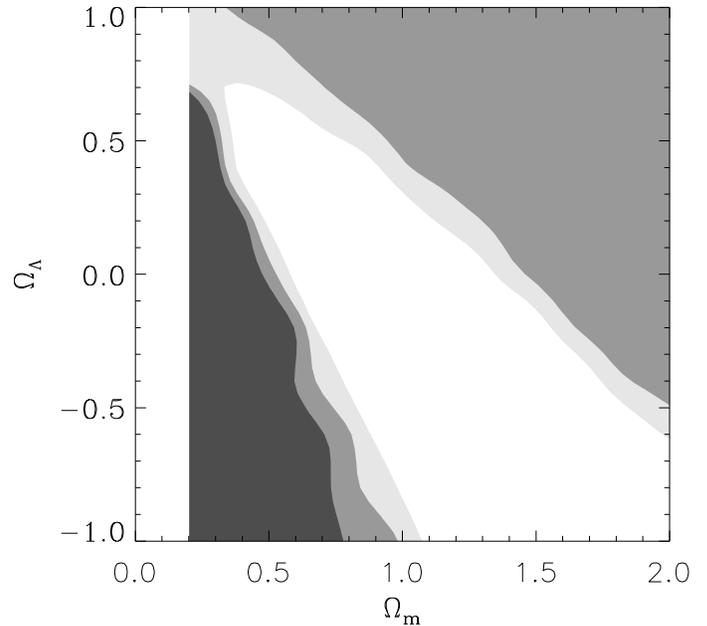}}
\caption{\label{boom2Dfig}\footnotesize%
Same as Figure~\ref{all65_2Dfig}, but
using only COBE and Boomerang data sets. The yellow/light grey contour
corresponds to the result of Melchiorri et al (2000) if we impose
$\on=0$.
}
\end{figure}

Because of the well-known angular diameter distance degeneracy, 
where increasing $\Ok$ shifts the acoustic peaks to the right and 
increasing $\Ol$ can shift them back to the left, we also plot
our constraints marginalized onto the 2-dimensional 
$(\Om,\Ol)$-plane.
\Fig{all65_2Dfig} shows the results using all the data, 
and Figures~\ref{lyman2Dfig}--~\ref{boom2Dfig} shown the constraints
from various subsets that will be described below.
In all cases, the shaded regions show what is ruled out at 
95\% confidence ($2-\sigma$). For our 2-dimensional parameter space, this
corresponds to $\Delta\chi^2=6.18$ (not 4), 
as in Press {\etal} (1992) \S 15.6. 

We show four nested contours. The least constraining one is when all 10
parameters are treated as unknown. The second includes our Hubble parameter
prior $h=0.65\pm 0.07$. The third (what we call our ``basic result'') 
adds the nucleosynthesis constraint
$\ob\approx 0.02$ and the fourth imposes $r=\tau=0$. Although the first two
priors are observationally well-motivated, the last one is completely ad hoc,
and has only been included to illustrate the importance of including 
reionization and gravity waves in analyses of this kind. 

When removing a prior constraint ($\ob=0.02$) 
from our basic result, we reduce all $\chi^2$-values by unity 
before plotting the corresponding contour, to account for the added 
degree of freedom. Similarly, we subtract 2 when dropping 
both constraints and add 2 when imposing $\tau=r=0$.

\Fig{all65_2Dfig} shows that the CMB data alone is able to rule
out very open ($\Om\simlt 0.4$) models with $\Ol=0$. 
Adding the $h$-constraint tightened the limits somewhat, mainly on 
very closed models. A more important prior at at this stage is 
that from nucleosynthesis, which helps eliminate most of the remaining 
closed models and places the first {\it lower} limit on $\Ol$.
This makes the allowed region in the $(\Om,\Ol)$-plane bounded, which
is important: otherwise all other constraints, which are marginalized over
$\Ol$, would depend sensitively on the poorly 
motivated prior $\Ol\ge -1$ that was hard-wired into our parameter grid.

Adding the additional prior $\tau=r=0$, which we recommend against for
the reasons described in the introduction, is seen to rule out
about half of the remaining models.
The exclusion of these parameters is seen to 
predominantly rule out closed models, whose 
first acoustic peak is too far to the left.
This is because it can be shifted back to the right by
tilting the power spectrum (increasing $\ns$), after which the
peak height can be brought back down to allowed levels 
using reionization or gravity waves.
In contrast, it is not possible to salvage too open
models with this trick: decreasing $\ns$ would 
require raising the first peak, but there is of course
no such thing as negative reionization or negative 
gravity waves.

\subsection{Is everything consistent?}

The plots we have shown so far are Bayesian in nature, and can only be
interpreted as advertised if the data are consistent with the best fit model.
This is indeed the case, since we obtain
$\chi^2 = 49$ for the best fit model. Dropping the
constraints on $h$ and $\ob$ reduces $\chi^2$ by as much as 6,
corresponding to the rather unphysical model shown in \fig{BestFitFig}.
For comparison, 65 data points and 10 parameters gives 55 degrees of 
freedom\footnote{In fact, our parameters do not span a full 10-dimensional 
subspace of the 65-dimensional data space when they range over physically 
reasonable values, since some of them have only a minor impact 
(say $\nt$) or are subject
to near degeneracies like $(\As,\tau)$, $(\Ok,\Ol)$ and $(\oc,\on)$.
The effective number of degrees of freedom to subtract off
may therefore be closer to 6 than 10.
}, so we should expect $\chi^2=55\pm 21$ at $2\sigma$.

\subsection{Robustness to choice of data}

To investigate the relative constraining power of different data sets and the degree
to which they give consistent results, we repeated our analysis for three subsets
of the observations. Specifically, we partitioned the most recent observations reporting 
multiple band powers into three disjoint sets and combined each one of them with 
the COBE measurements:
\begin{enumerate}
\item The ``East Coast'' sample contains Saskatoon, QMAP, TOCO and COBE.
\item The ``snake'' sample contains Python, Viper and COBE.
\item The Boomerang sample contains Boomerang-97 and COBE.
\end{enumerate}
As seen in 
Figures~\ref{lyman2Dfig}--\ref{boom2Dfig}, they all allow flat models and 
disfavor very open ($\Ok\gg 0$) models, 
which would place the first acoustic peak too far to the right. 
As more priors get added, they are seen to disfavor very closed models as well.
In all cases, the best fit model has an acceptable $\chi^2$-value.

\subsection{Importance of calibration errors}

To assess the importance of calibration uncertainties, 
we repeated our analysis with all calibration errors set to zero.
We found that in this case, {\it no} model provided a very good fit 
to the data, with $\chi^2\approx 76$ for the best fit.
This is only a problem at the $2\sigma$-level
for 65-10=55 degrees of freedom, and perhaps even less in light
of footnote 1.
However, it nonetheless caused the the Baysean constraints 
to become quite misleading, suggesting that 
most parameters were very tightly constrained around
their maximum-likelihood values --- for instance, that $\on=0$ was ruled out 
at high significance. In conclusion, it is of paramount importance to include 
calibration errors. This was done in the above-mentioned analyses 
of Dodelson \& Knox (2000) and Melchiorri {\etal} (2000), 
but not in most earlier work.

The main discrepancies pushing up the $\chi^2$ were localized to two places in \fig{DataFig}.
The first trouble spot was at $40\simlt\ell\simlt 70$, where
the low Python V points conflicted with the
higher measurements on a similar scale from, \eg, Toco, QMAP and Saskatoon.
The second problem occurred at $\ell\simlt 300$, where the models
failed to fall rapidly enough from the high Toco detections down to the
lower power levels seen by MSAM, CAT, OVRO, Viper
and Boomerang.

\subsection{Are the data internally consistent?}

\begin{figure}[tb] 
\centerline{\epsfxsize=9.7cm\epsffile{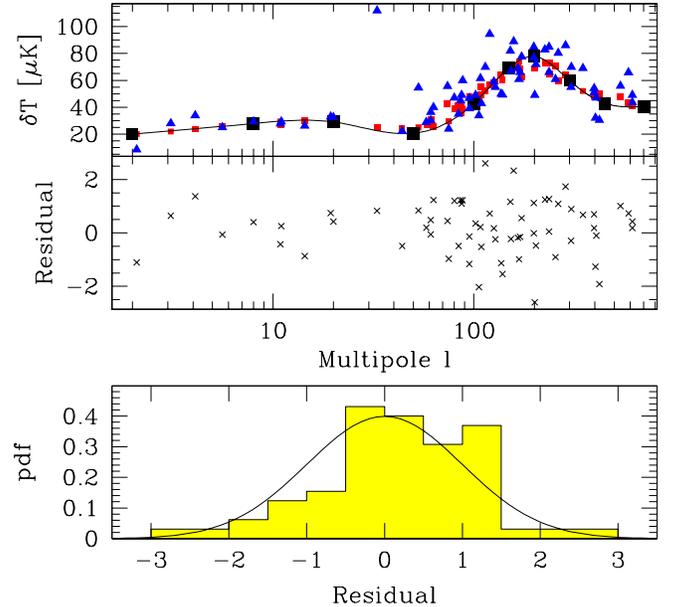}}
\vskip-1.0cm
\caption{\label{ResidualFig}\footnotesize%
Residuals. 
The top panel shows a cubic spline interpolated between 10 equispaced 
grid points (large squares) that are adjusted vertically to 
make the window function convolved curve (small squares)
fit the observations (triangles) as well as possible.
The middle panel shows the residuals 
$(d_i-\expec{d_i})/\sigma_i$, the differences between the 
triangles and small squares in units of the error bars. The bottom panel shows a histogram of these residuals
compared with a unit Gaussian. The reduced $\chi^2$-value is simply
the second moment of this distribution.
}
\end{figure}

Based on visual inspection of the data, it has been suggested
that all CMB measurements cannot be consistent with 
any model, since some measurements disagree with others on a comparable
angular scale. Although we saw above that the $\chi^2$-value
is acceptable, the distribution of residuals could in principle be non-Gaussian 
with the a few severe outliers being averaged down beyond recognition in the
$\chi^2$-calculation.
To investigate this possibility, we fit a 10-parameter model with no 
underlying physical model to the data. Our model curve 
is simply a cubic spline interpolated between 10  
grid points. 
\Fig{ResidualFig} shows the 65 residuals $(d_i-\expec{d_i})/\sigma_i$,
ignoring calibration errors, and reveals no striking outliers at all.

This fit gives a $\chi^2\approx 67$ ignoring calibration errors, \ie, 
even lower than for the CMB case.
In view of footnote 1, we repeated this test with merely six spline points.
This gave $\chi^2\approx 95$ for logarithmically equispaced spline points,
but as low a $\chi^2$ as before when more points were shifted to be near the
1st acoustic peak.

In the future, as CMB data gets still better, one would expect the correct
physically based model to provide a substantially better 
fit than ``any old smooth curve'' with the same number of free parameters.
Until then, \ie, until our physical theory provides the most economical
explanation of the observations, 
we cannot interpret the good fit of the model to the data as 
overwhelming evidence that our theory is correct.

\section{CONCLUSIONS}

\label{ConclusionsSec}

\begin{figure}[tb] 
\centerline{\hglue6.0cm\epsfxsize=15.0cm\epsffile{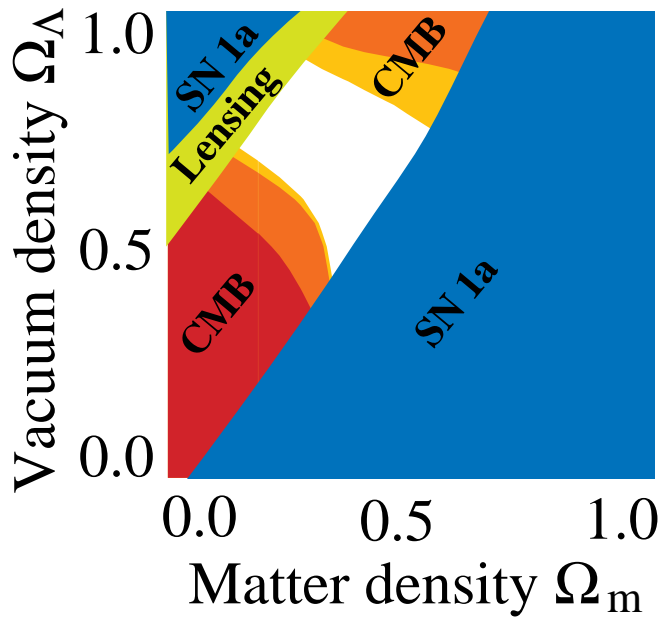}}
\vskip-4.5cm
\caption{\label{EverythingFig}\footnotesize%
Constraints in the $\Om-\Ol$ plane.
The regions in the $(\Om,\Ol)$-plane 
that are ruled out by our analysis at 
$2\sigma$ using all the data are shown 
using no priors (red/dark grey), 
the prior $h=0.65\pm 0.07$ (orange/grey), and the
additional nucleosynthesis constraint $h^2\Ob=0.02$ 
(light orange/light grey).
The SN 1a constraints are from White (1998) and
the lensing constraints are from Falco et al (1998).
The CMB contours for $\Om<0.2$ are extrapolations.
}
\end{figure}

We have presented a method for rapid calculation of large numbers of CMB models
and used it to jointly constrain 10 cosmological parameters from current 
CMB data. Our results on individual parameters are summarized in Table 2. 
Arguably the most interesting constraints at this point are those on the
geometry of spacetime, summarized in \fig{EverythingFig}.
This figure zooms in on the upper left quarter of \Fig{all65_2Dfig}
and shows the joint constraints on $\Om$ and $\Ol$ from a variety of
astrophysical observations. The SN 1a constraints are from White 1998,
combining the data from both search teams 
(Perlmutter {\etal} 1998; Riess {\etal} 1998; Garnavich {\etal} 1998).
As can be seen, the CMB and SN 1a constraints imply a positive 
cosmological constant ($\Ol>0$) when combined. If the 
Falco {\etal} (1998) constraints from 
gravitational lens statistics are included, the allowed region
in parameter space is further reduced.

This claim that $\Ol>0$ is of course old hat (Kamionkowski \& Buchalter 2000), 
originally being made over a year ago
(see Sahni \& Starobinsky 2000 for a recent review).
What is new here,
and quite striking, is its robustness. Since the first such
joint analysis (White 1998), the number of CMB band power
measurements has roughly doubled, with experiments such 
as Toco, Python V, Viper and Boomerang greatly improving the
accuracy on acoustic peak scales. In addition, the CMB 
treatments has been gradually refined; for example, several
groups have added calibration errors and this analysis has
weakened the constraints further by fitting for 10 parameters
jointly. Yet despite these major improvements in both data and modeling,
the cosmological constant remains alive and well, stubbornly refusing
to vanish.

Since CMB data is likely to continue to improve at a
rapid pace, with exciting new balloon, interferometer and satellite data
just around the corner, it will be important to further improve on 
the type of analysis that we have presented here.
There are a number of areas in which the accuracy off our treatment 
can be improved:
\begin{itemize}

\item The problem of regenerated power from very early reionization
can be eliminated by explicitly looping over $\tau$ 
for the high $\l$ models in Step 1
instead of using the $e^{-2\tau}$ suppression approximation.
 
\item In Steps 1 and 2, the effect of gravitational lensing can be included.

\item In Steps 1 and 2, further speed-up can be attained by taking advantage
of the fact that the tensor fluctuations are essentially 
independent of $\on$ as long as 
the total dark matter density $\oc+\on$ stays constant. 

\item The accuracy in Step 2 can probably be further improved by 
using some form of morphing technique as suggested by 
Sigurdson \& Scott (2000).
The basic idea is to interpolate not the power spectrum itself
but some cleverly chosen parametrization thereof. We have done this to 
a certain extent by computing and interpolating the amount by which
the acoustic peaks should be shifted sideways, but more ambitious
reparametrizations are clearly possible.

\item In Step 3, the likelihoods can be computed more accurately
by incorporating non-Gaussianity corrections as in
Bond, Jaffe \& Knox (1998) or Bartlett {\etal} (1999)
and by including correlations between different 
data points. The former is particularly important for 
upper limits, which were simply excluded from the
present analysus. The latter includes correlations between 
different experiments that overlap in sky coverage and angular scale.
Calibrations can be treated as multiplicative 
parameters to be marginalized over (as in Dodelson \& Knox 2000) rather
than as correlated noise (our approximation is accurate as
long as the relative calibration errors are much less than unity).

\item Step 3 should ideally use the exact band power window functions.
Unfortunately, most window functions available in the literature 
are variance window functions, and using them as band-power window 
functions is an approximation which is not
always good. Experimentalists are strongly encouraged
to publish their band power window functions!

\item The overall accuracy of our technique can be improved
with brute force, by computing a finer grid of models in step 1.
Indeed, the errors introduced in Step 2 can in principle
be continuously reduced toward zero by
refining the $(\oc,\ob,\on)$-grid for low 
$\l$ and shifting the splicing point upwards from $\l\sim 100$.

\item The accuracy in Step 4 can be improved
by integrating instead of maximizing when marginalizing.
This will make a difference mainly early on when the 10-dimensional 
probability distribution in parameter space is widely extended and
differs greatly from a multivariate Gaussian distribution.
If this integration approach is used, it should be applied
even for the normalizations $\As$ and $\At$, for consistency.

\end{itemize}

A second general area of improvement will be to include more prior 
information than Hubble parameter measurements and 
nucleosynthesis constraints.
As data improves in a wide variety of areas, 
this will not only help break parameter degeneracies, but also
allow important cross-checks.
A very large number of such multi-dataset studies have been carried 
out in the past (Bahcall {\etal} 1999 and Bridle {\etal} 1999 
provide good recent entry points into the literature), 
but rarely for more than a few parameters at a time.
Here is a necessarily incomplete list
of such constraints: 

\begin{itemize}

\item Measurements of the matter power spectrum and its time-evolution 
$P(k,z)$ from galaxy redshift surveys.

\item
Measurements of $P(k,z)$ from weak gravitational lensing
(\eg, Narayan \& Bartelmann 1996)

\item 
Measurements of $P(k,z)$ from
the abundance of galaxy clusters 
(\eg, Carlberg 1997; Bahcall \& Fan 1998; Eke {\etal} 1998.)

\item
Constraints om $P(k)$ from peculiar velocity measurements
(\eg, Zehavi \& Dekel 1999).

\item Limits on $(h, \Ok, \Ol)$ from SN 1a.

\item Limits on $(h, \Ok, \Ol)$ from lens statistics
(\eg, Kochanek 1996; Falco {\etal} 1998; Bartelmann {\etal} 1998;
Helbig 1999)

\item Limits on $(h, \Ok, \Ol)$ from 
limits on the age of the Universe and various other 
classical cosmological tests (Peebles 1993).
For instance SZ cluster distance measurements provide promising
new constraints of this type (Reese {\etal} 2000).

\item Direct measurements of $\Om$ and 
the baryon fraction $\Ob/\Om$ from cluster studies
(Carlberg {\etal} 1999; White {\etal} 1993; Danos \& Pen 1998; Cooray 1998) 

\end{itemize}
    
Finally, adding more physics can both weaken and tighten constraints.
Adding further parameters (say an equation of state for a scalar field 
component) can weaken constraints on other semi-degenerate
parameters. On the other hand, 
adding an astrophysical model for, say, how $\tau$ 
depends on the other parameters can substantially
tighten constraints (Venkatesian 2000).

In conclusion, as CMB experimentalists continue to 
forge ahead, CMB theorists will need to work hard to keep up.


\bigskip
The authors wish to thank Lloyd Knox for helpful comments about
calibration errors
and Ang\'elica de Oliveira-Costa, 
Mark Devlin, Charley Lineweaver and Amber Miller for useful 
discussions.
Support for this work was provided by NSF grant AST00-71213, 
NASA grant NAG5-9194,
the University of Pennsylvania Research Foundation, and 
Hubble Fellowships HF-01084.01-96A and HF-01116.01-98A from 
STScI, operated by AURA, Inc. 
under NASA contract NAS5-26555. 

\appendix

\section{Conditional marginalization}

\def\fmax{f_{\rm max}}
\def\fint{f_{\rm int}}
\def\x{{\bf x}}
\def\y{{\bf y}}
\def\z{{\bf z}}
\def\xb{\bar\x}
\def\yb{\bar\y}
\def\zb{\bar\z}
\def\D{{\bf D}}
\def\E{{\bf E}}
\def\F{{\bf F}}
\def\G{{\bf G}}
\def\I{{\bf I}}
\def\J{{\bf J}}
\def\K{{\bf K}}
\def\L{{\bf L}}
\def\M{{\bf M}}

In this Appendix, we show that maximizing is equivalent to integrating 
when marginalizing multidimensional Gaussians in arbitrary dimensions. 
Although this useful property is undoubtedly derived in the statistics
literature, we present a brief derivation here for completeness.  

A multivariate Gaussian distribution in $n$ dimensions takes the form
\beq{GaussEq}
f(\x) = (2\pi |\C|)^{-n/2} e^{-{1\over 2}(\x-\xb)^t\C^{-1}(\x-\xb)},
\eeq
where $\xb$ is the mean vector and $\C$ is the $n\times n$ covariance matrix.
Let us partition the $n$ parameters in $\x$ into two subsets $\y$ and $\z$
of size $n_x$ and $n_y$ ($n_x+n_y=n$) and write 
\beq{DecompEq}
\x\equiv\left(\y\atop\z\right),
\quad
\C^{-1}=\left(\begin{tabular}{cc}
$\D$&$\E$\\
$\E^t$&$\F$
\end{tabular}\right).
\eeq
We can now define a probability distribution 
for $\y$ in two different ways, by either integrating or maximizing over $\z$:
\beq{fintDefEq}
\fint(\y)\equiv\int f(\x)d^{n_z} z,
\eeq
\beq{fmaxDefEq}
\fmax(\y)\equiv c \max_{\z} f(\x),
\eeq
where the normalization constant $c$ is chosen so that $\fmax$ 
integrates to unity.
Maximizing $f$ is equivalent to minimizing the quadratic
form $(\x-\xb)^t\C^{-1}(\x-\xb)$. Inserting \eq{DecompEq} and differentiating with
respect to $\z$ shows that this minimum is attained for 
\beq{optzEq}
\z = \zb - \F^{-1}\E^t(\y-\yb).
\eeq
Substituting this back into \eq{fmaxDefEq} gives
\beq{fmaxEq}
\fmax(\y) \propto e^{-{1\over 2}(\y-\yb)^t[\D-\E\F\E^t](\y-\yb)},
\eeq
\ie, a Gaussian with mean $\yb$ and covariance matrix $[\D-\E\F\E^t]^{-1}$.
As is well known, integrating over $\z$ also gives a Gaussian with
mean $\y$ and a covariance matrix which is simply the upper left
submatrix of the full covariance matrix $\C$.
The identity
\beqa{MonsterEq}
&&\hskip-0.7cm\left(\begin{tabular}{cc}
$\D$&$\E$\\
$\E^t$&$\F$
\end{tabular}\right)^{-1}
= \\
&&\hskip-0.7cm\left(\begin{tabular}{cc}
$[\D-\E\F^{-1}\E^t]^{-1}$&$-\D^{-1}\E[\F-\E\D^{-1}\E^t]^{-1}$\\
$-\F^{-1}\E^t[\D-\E\F^{-1}\E^t]^{-1}$&$[\F-\E\D^{-1}\E^t]^{-1}$
\end{tabular}\right)
\nonumber
\eeqa
therefore shows that the covariance matrix is $[\D-\E\F^{-1}\E^t]^{-1}$,
\ie, the same as for the maximization case.
This proves that $\fint=\fmax$, \ie, that the two methods of marginalization
give identical results when the probability distribution is Gaussian.
The identity given by \eq{MonsterEq} is readily proven by simply
multiplying the matrices on the left and right hand sides together
and verifying that their product is the identity matrix.



\end{document}